# SOA-based security governance middleware

Pierre de Leusse[*1] and Theo Dimitrakos[**]
[*]AGH University, [**]BT Innovate & Design
pdl@agh.edu.pl

## Abstract

*Abstract*—Business requirements for rapid operational efficiency, customer responsiveness as well as rapid adaptability are actively driving the need for ever increasing communication and integration capabilities of software assets. In this context, security, although acknowledged as being a necessity, is often perceived as a hindrance. Indeed, dynamic environments require flexible and understandable security that can be customized, adapted and reconfigured dynamically to face changing requirements. In this paper, the authors propose SOA-based security governance middleware that handles security requirements on behalf of a resource exposed through it. The middleware aims at providing different security settings through the use of managed compositions of security services called profiles. The main added value of this work compared to existing handlers or centralized approaches lies in its enhanced flexibility and transparency.

*Keywords-SOA, SOA governance, SOA security, SOA adaptability*

## I. INTRODUCTION

The way enterprises conduct their business is constantly changing and the need for adaptation and seizure of emerging opportunities is often strong. Enterprises have become more pervasive, with mobile workforces, outsourced data centers, different engagements with customers, suppliers and distributed sites [1]. This increases the need for securing end-to-end transactions between business partners and customers. Yet, as presented in [2], fully or correctly incorporating security is often seen as an impediment to the process of writing and deploying software.

In order to achieve agility and shorter concept-to-market timescales for new products and services, ICT service providers and their corporate customers alike increasingly adopt a collection of technologies, concepts and capabilities which come under the banner of Service Oriented Architecture (SOA). The Security as a Service (SaaS) paradigm can be a means of providing security-related functionalities in a SOA context.

The aim of this paper is to show how a governed composition of security-related services (e.g. access control, identity management), provided through the SaaS paradigm, can be leveraged upon in order to provide a flexible and auditable approach to security in distributed and complex systems. This paper features security governance middleware that allows manipulating the security configuration of exposed resources in a more dynamic and flexible way compared to existing techniques such as handlers [3] or the Security Service Bus [4]. An additional key aspect of this security governance middleware is to improve the visibility of various parameters taken into account when securing access to resources in order to facilitate the decision making process. Throughout this paper, the authors attempt to demonstrate the practicability of governed, composable and adaptable security.

In section II related work is outlined and its drawbacks discussed in the presented context. In section III the SOA governance architecture developed by the authors is introduced. In section IV a specific use case of this architecture, dedicated to managing security is presented through a virtual music store scenario. Finally an evaluation of this security-oriented use case is made and supplemented by conclusions. The paper should be considered a more detailed version of [5], where the authors briefly outline their work.

## II. RELATED WORK

In the security area, projects like GOLD [6] and TrustCoM [7] have demonstrated the feasibility of policy-based security solutions in decentralized environments. These solutions, however, present only limited capacity to adapt to changes in policy (e.g. grammar) or its enforcement processes (e.g. when a threat is discovered).

In [8] and [4] flexible security policy enforcement architectures are proposed. However, they do not provide the same level of dynamicity and evolution, as they require more human involvement (e.g. administrator, developer) in order to add new security services or take new security contracts into account.

---

[1] Pierre has only recently moved to AGH University and was previously working at Newcastle University.

### III. ANATOMY OF GOVERNANCE MIDDLEWARE

The goal of this governance middleware is to manage the exposure of a resource over a network. In the context of this paper, such a client resource is called Business Capability. This capability is achieved by managing properties of the exposure, including its non-functional properties (NFPs) (e.g. audit, transport protocol) as well as the lifecycle of the infrastructure put in place to govern the exposure and the services enacting NFPs.

For the presented governance to be effective, each of these Infrastructure Capabilities (i.e. a service that enacts an NFP) is expected to be deployed as a web service with its own service management, policy administration framework (control pane) and operational interfaces (data pane). Each capability is also expected to be policy-driven as this permits configurability and flexibility. Different capabilities are each meant to have their own distinct grammar and policy languages in order to maintain their own advantages, capacities and evolution potential in their respective domains. This approach enables the interchange of core capabilities within respective categories, according to the NFP requirements (whenever necessary).

In order to achieve this, the NFP requirements of the exposed resource are expressed in a normalized manner as a profile which is used to define the way the resource is to be exposed.

The architecture is divided into three distinct dimensions: operational, data and management. The operational dimension defines a set of core elements forming the execution of the governance middleware itself (e.g. access control, identity brokerage, service registry). The data layer allows describing various aspects of governance. Finally, the management layer supports the interactions between different elements of the infrastructure.

This governance middleware provides a management interface as an entry point for both selecting the profile type (if one has been defined) and specifying the context in which a profile type will operate.

More details on this architecture can be found in [9, 10]. In the following sections, an implementation of this framework dedicated to governing the security aspects of web services exposure and interactions is presented.

### IV. VIRTUAL MUSIC STORE SCENARIO

#### A. Description

This section describes the scenario of a Virtual Music Store (VMS) as an example of a Virtualized Organization (VO). VOs can be loosely defined as temporal collaborations between organizational entities [11]. According to [12], VOs are frequently restructured, sustained to capture the value of a market opportunity and dissolved again to pave the way for a new VO created from within a network of independent partners. As such, VOs represent an opportunity for adaptability that current systems, such as the GOLD middleware [6] or the TrustCOM B2B gateway [7, 13] also attempt to address.

The presented security governance middleware intercepts messages addressed to a VMS, enforces security policies and integrates the defined security capabilities (e.g. identity management, authorization service) in order to secure VMS communications with its suppliers.

With this scenario, the authors aim to demonstrate that the architecture introduced in section 3 can be applied to provide complex NFPs such as access control, identity management and policy enforcement. Indeed, these security-related NFPs necessitate configuration of the capabilities both individually and collectively. For instance, in a certain federation anatomy, a Secure Token Service (STS) providing identity management will not only have to be configured, but will also require a means for establishing trust with the different STSs it is meant to interact with. The situation can be different in other federation anatomies, thus necessitating tight management of how to govern the assembly of capabilities in a safe manner.

#### B. Partners and roles

In this scenario, the aggregated services are virtual music stores serving specialized markets or communities of interest. The basic service providers include copyright owners of musical recordings or their representatives who make such recordings available online. The music stores reach agreements with music providers enabling them to act as re-sellers of bundles of recordings from their respective catalogues. The VMS is a VO consisting of the music store operator communicating with content providers through security governance middleware whose security services are provided by external security providers using the SaaS paradigm.

The end customer of the VMS will be a member of the public. What they will see is a web site where they are able to search for and buy tracks, and access music-related content (e.g. reviews, blogs). This content could be presented to them in much the same way as AbeBooks does, i.e. a search page, with each returned item linked in by an independent seller. Alternatively, stores could hide the aggregated nature of the service.

As shown in Figure 1, in the music store scenario, the main partner categories are:

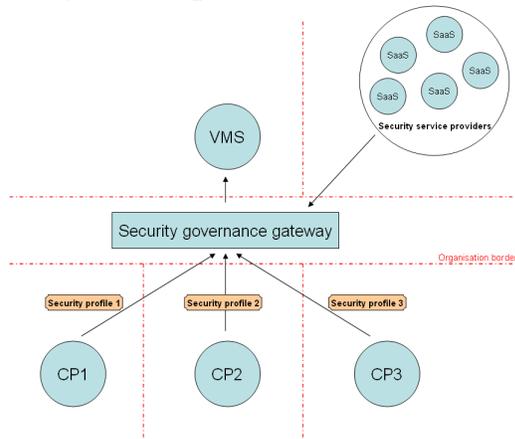

Figure 1. Virtual music store scenario with security governance middleware

**Content Provider (CP)**: this is a specialized content provider (e.g. record labels or other copyright owners). Three different content providers take part in this particular scenario. Each uses a different security profile which it shares in a secure way with the VMS operator through the security governance middleware. The security profiles express formalized security requirements in a set of XML files, as illustrated in Figure 1. The following is a list of content providers, with an overview of their security settings:

- CP1 requires HTTP basic access authentication.
- CP2 requires SecPAL to express its access control requests and identify the requesters.
- CP3 requires login and password with its custom XML token schema.

**Security governance provider**: this role involves providing the VMS with security governance middleware. The purpose of such middleware is to hide the technical complexity of the security system involved and enhance the visibility of security actions that need to be enforced. In addition, it allows connecting the VMS to different content providers that expose their catalogs using diverse security configurations.

**Virtual Music Store (VMS) operator**: this is the music broker. Ultimately, the administrator of the VMS is responsible for selecting its CP partners.

**Security a a Service (SaaS) provider**: this is a third party entrusted with providing a security service such as identity management, access control or audit. These services are managed by security governance and allow the content providers and music store operator to leverage the security governance middlware to enhance their interoperability while preserving the necessary level of security. In this experiment, the services shown in Figure 4 are registered and described using the same taxonomy as the one used in the security profile. Although these providers are not known from the VMS and CPs, for the sake of this experiment it is assumed that actors trust the security governance provider who, in turn, trusts the security service providers. The authors believe that in a production environment this issue would be overcome by providing solutions such as QoS or dynamic trust-based selection of services.

*C. Security governance lifecycle*

This section describes the lifecycle of the virtual music store security. It starts with the governance middlware configuration, then carries on with initial agreements and discovery of potential partners, leading to the formation of a new VO. This is followed by security profile management, facilitating verification and instantiation of the security profile supplied by content providers. Finally, the adaptability faculty of the virtual store infrastructure is introduced.

**Security governance middleware configuration**. Prior to forming the VO and at any time during the following steps, the security governance middleware is meant to reach agreements with security service providers to use their services within a particular context (e.g. VO) and/or with specific requirements (e.g. security settings) in order to provide security for the different content providers used by the VMS. In a production environment, these agreements could be reached using human or electronic negotiation protocols.

**VO formation**. Prior to any collaborative task and once the virtual shop has decided to establish the music store, it needs to reach an agreement with the security governance provider.

With a governance middleware in place, the VMS operator contacts the potential participants of the music shop (i.e. CPs). Agreements are reached between these content providers and the VMS operator regarding the conditions in which the business will be conducted and the VO will operate. Subsequently, the CPs deposit their security profiles in the middleware together with the relevant data about their business functions (e.g. music catalog). These security profiles contain the data necessary for the governance middleware to understand how to securely connect to CP services. More specifically, this means that CP3's security profile will comport a reference to the appropriate login XML schema. CP2's security profile, on the other hand, will provide (for instance) a reference to the SecPAL assertion it expects along which information on which service, interface and operation the security governance middleware needs to contact in order to establish trust and therefore allow for its identity to be established.

**Security profile management**. At this stage it becomes possible for the security middleware to create the different security profiles.

To achieve this, the profile management component of the middleware goes through a set of actions that starts with the verification of the security profile proposed, then defines the processes necessary for the instantiation and management of the future implementation of the security profile.

Profile management is divided into two main logical domains: profile consistency management and profile lifecycle management. Each of these domains can be further split into several steps: defining the infrastructure capabilities, policy templates, service dependencies and information flow (for consistency management) and defining the profile management process and publishing the infrastructure profile (for lifecycle management). The following steps detail the security profile management lifecycle.

1. Define service description;
2. Define service policy scheme;
3. Define service usage policy;
4. Define service management process.

The first stage of this process is to check, for each security service, whether the description given in the security profile is adequate and complete according to the profile description taxonomy. If that test passes, the abstract definition of the service as well as its requirements are gathered. At this stage the governance has not selected a particular instance of any running service but has selected the appropriate categories in the domain description taxonomy.

The security governance middleware separates copies of the security profile at the beginning and end of this process.

5. Select service;
6. Define policy template;
7. Define domain of metadata transformations (I/O metadata);
8. Define policy management processes.

This second stage verifies if the policy type that will be used requires any translation (when using different grammar types). In addition, this stage checks how they need to be managed.

9. Select security service;
10. Define operation bindings;
11. Define security service invocation pattern;
12. Validate security service dependencies.

Following this individual check, the relationships between the services, both on the managerial and operational sides, are verified.

The last stage involves going through the security profile and checking for missing components. For instance, a content provider could have specified an access control method without mentioning any identity management.

Following this, the VMS operator can review and select the best matches from among the positive answers it has received, potentially eliminating content providers with weak or costly security configurations.

13. Select security services;
14. Define policy metadata transformations;
15. Validate policy dependencies.

With the security profile complete and safe, the fourth stage aims at checking how the data will flow from one operation to the next and determining, whenever relevant, whether a translation process is necessary amongst them and/or with the content provider's security settings. Please note that for simplicity's sake this experiment does not focus on or implement the translation itself.

The result of this step is saved as a security profile that can be instantiated.

16. Select security service management processes;
17. Select policy management processes;
18. Define coordination process;
19. Bind management processes with coordination process;
20. Validate dependencies.

With the profile in place and the services selected, the final stage consists in the profile management component defining the different steps that will be necessary in order to call, configure and connect the security services.

The result of this last stage is stored together with the now-complete safe security profile as a profile management process.

When necessary, the security governance middleware can enact this complete profile through its management process and, upon the completion of this enactment, begin exchanging messages with each content provider. Moving from a security profile (as provided by steps 1-20) to its enacted version involves the following:

21. Discover infrastructure profiles (exposure of context);
22. Select infrastructure profile(s);
23. Define bindings to business capability;
24. Validate service dependencies.

When required to enact a security profile, the middleware searches its database for a suitable match. It then makes sure that data can flow between the enacted security profile and the VMS connections to the secured CP services.

25. Select infrastructure capability;
26. Refine policy template-specific policies;
27. Update capability policies.

Profile management iterates through all the specific services selected in the security profile and instantiates policy templates by inserting the context data.

28. Select infrastructure capabilities;
29. Refine policy metadata transformations (service-specific metadata);
30. Validate policy dependencies.

If necessary, the transformations templates are also instantiated according to the results of step 14.

31. Select profile management processes;
32. Select business capability management processes;
33. Define coordination process;
34. Bind management processes with coordination process;
35. Validate dependencies.

The profile manager orders and binds the different management processes of the security profile and the VMS connections to the CP services they protect.

36. Update capability policy stores;
37. Update infrastructure bindings;
38. Expose business capability to service endpoint;
39. Publish service.

Finally, the policies created in step 26 are pushed to the relevant stores (e.g. the SaaS capabilities driven by them) through the enactment of management processes. Finally, the VMS clients of the CP services are bound to their instantiated profiles and exposed.

**Adaptability**. A VMS will want to be able to include as many content providers as it can, but different partners will have distinct security needs and settings. By reviewing and accepting the different security profiles used by each partner according to its specific needs, the VMS can more promptly make use of their contents. This both necessitates and is limited by the capability of the security governance middleware to find security-related services that provide for these different requirements.

## V. EVALUATION

The purpose of this evaluation is to demonstrate the adaptability of security governance middleware. In order to do so, the authors have assessed its capacity for integration with different systems and the capacity to manage this adaptability when dealing with different types of unavailability. Communications between the VMS and the security governance solution are secured using the VMS security settings. The security of this type of middleware is assumed to be equivalent to that of the security profiles it enacts. Indeed, the message exchanges between the different components of the middleware are secured as well as the exchanges with external services. External security services that are used to provide security requirements are trusted.

In [4] (cf. related work section), the authors have devised six types of adaptation that a flexible security middleware suite should support:

**S1** Change a local parameter of a security component (e.g. the encryption method for an audit service).

**S2** Introduce new security functionality (e.g. add a secure logging component).

**S3** Compose/recompose a deployed security component with one or more application components. Application components depend on the security component but the security component can also depend on the application component(s) (e.g. for context-based access control).

**S4** Swap a security component for another one (e.g. replace the authorization decision engine).

**S5** Compose a security component using a (new) third-party component that is deployed elsewhere.

**S6** Change a security policy. Since the security policy explicitly depends on application-level concepts, any change in a security policy can require further adaptations.

The system proposed in this paper, based on the security profile, is meant to introduce flexibility to the way in which the NFP aggregation lifecycle is managed and user requirements expressed. Therefore the following evaluation points have been added:

**S7** Enact the security profile at different stages in different situations (e.g. CP1 steps 1-39, CP2 steps 21-39).

**S8** Express identical requirements using different semantics.

### A. Qualitative evaluation

The objective of this type of evaluation is to ensure that the adaptability functions as intended and to define its limits. The security profiles submitted by the content provider to the security governance middleware have been used in order to determine the scope of the adaptation.

**S1** Changing a security service's configuration requires the user to change the security profile. With the change committed, the governance middleware will make use of Service Management's access to the security service's management interface to perform the change. However, the governance infrastructure will go through the process of profile management to ensure that the change is valid and can be realized. If the security service does not support this change a different one may be selected or the modification rejected.

**S2** If the relevant security services are registered as accessible in this context and with these requirements, a security profile needs to be updated with the additional security functionality.

**S3** Composition of security as well as other value-adding services can be realized. Of course, the quality of the segregation of a security service between different contexts (e.g. different interactions and conversations) depends on its implementation and may not be possible. For instance, a service may or may not support multi-domain instantiations and configurations.

**S4** As previously presented, replacing a security service with another, similar one is achievable as long as a potential replacement is registered and accessible. If a security service is found missing at runtime, it is possible to start again from step 13 onwards, to regenerate an instantiated profile while storing the incoming and outgoing messages.

**S5** The ability to compose external providers' security services is the very foundation of this work.

**S6** In this model, this is equivalent to **S1**. The potential transformation required to interact with other security services will also be adapted.

**S7** It is possible to store a profile's state at any stage of its lifecycle. However, validation stages are required in most cases to ensure the profile's validity at the time of use (e.g. whether the security service is still available).

**S8** This point has not been verified in this set of experiments. However, it is defined in [9] and depends on the different semantic styles used as well as their interoperability.

## VI. CONCLUSIONS

In this paper, the authors provide an overview of an architecture for SOA governance along with a scenario instance dedicated to managing the security aspects of message exchanges between distributed resources. This instance has been evaluated against a virtual music store scenario where it is necessary to adapt to three different security settings. An evaluation of how the security governance middleware provides adaptable security in regards to this scenario is proposed.

It is noticeable that the increased complexity of the governance middleware, compared to the more traditional and static approach of handlers, implies an additional level of complexity. However, we believe this tradeoff is compensated for by the gains in flexibility and dynamicity, as well as by supporting per-action result audits (if necessary).

Future work will focus on investigating the possibility for the governance infrastructure to behave in a more autonomous manner to achieve self-* systems (self-adaptive, self-healing, self-growing etc.) For instance, when no similar service can be discovered, the content provider and the VMS could be advised to modify their security settings. The security governance middleware could then attempt to automatically find a relevant compromise, while waiting for the situation to be resolved.

Resource permitting, another direction for future work is the analysis of the architecture's scalability in a more complex scenario.